\documentclass[a4paper,fleqnb]{cas-sc}
\usepackage{hyperref}
\usepackage[numbers]{natbib}
\begin{document}

\shorttitle{Ferroelectric response to interlayer shifting and rotations in trilayer hexagonal Boron Nitride}    \shortauthors{E. Cortés Estay, J. M. Florez, E. Suárez Morell}  

\title[mode = title]{Ferroelectric response to interlayer shifting and rotations in trilayer hexagonal Boron Nitride}

\author[1]{Emilio Cortés Estay}[orcid=0000-0002-0736-8188]

\author[1]{Juan Manuel Florez}[orcid=0000-0002-4990-767X]
\ead{juanmanuel.florez@usm.cl}
\cormark[1]

\author[1,2]{Eric Suárez Morell}[orcid=0000-0001-7211-2261]
\ead{eric.suarez@usm.cl}

\affiliation[1]{organization={Grupo de Simulaciones, Departamento de Física, Universidad Técnica Federico Santa María}, 
adressline={Avenida España 1680, Casilla 110-V}, 
city={Valparaíso},
country={Chile}}

\affiliation[2]{organization={Departamento de Física Aplicada I, Escuela Politécnica Superior, Universidad de Sevilla}, 
postcode={41011}, 
city={Seville},
country={Spain}}
\cortext[1]{Corresponding author}

\begin{abstract}
From monolayers composed of different atoms, we can build structures with spontaneous vertical polarization by conveniently stacking multiple layers. We have studied, using first-principles methods and based on modern polarization theory, a system composed of three layers of hexagonal Boron-Nitride(h-BN) in all possible stackings. We obtain for each of the configurations how charge transfers between the layers and how it impacts on the polarization of the system. In addition, we studied a system of three layers, one of them rotated, and we found that not only did the magnitude of the polarization increase comparing with the bilayer but also, depending on the initial stacking and the rotated layer, we can create a variety of mosaic-like polarization arrangements, which are composed of regions with either triangular or hexagonal symmetry. 
\end{abstract}


\maketitle

\section{Introduction}
The advent of 2D materials has changed the way we design electronic devices. Dozens of 2D materials have been synthesized in the laboratory with different properties and with different spatial symmetries\cite{Geim2013,2DNovo}. It is also possible to couple atomic-thick monolayers of different or the same material conforming bilayers or multilayers with the desired stacking and even control the relative rotation angles between the layers.

Polar crystals have magnetic or/and electric polarization, with the electric order parameter of proper or improper nature. In the first case, the electrical parameter is mainly adjusted by applying an electric field or a voltage due to a strain-based pressure. In the second case, either a magnetic or electric field could be used to drive such order parameter depending on the magnetoelectric coupling, which could determine the appearance of one order due to the variation of the other. The so-called multiferroic materials are still scarce, and most of them work at low temperatures. 
Recently, progress has been made in the field of 2D oxides, and perovskites with few atomic layers have been synthesized. However, electric polarization has several limitations, and the ferroelectricity induced by oxygen stoichiometry seems too sensitive to various factors \cite{2dmagnetoelecUlrich2022,UltrathinferroYunseok2021,ProgressGuan:2020bca,LimitNordlander:2019hwa,emilio2022,Opazo:bm}. 

Finding a 2D material with spontaneous and robust ferroelectricity whose order parameter can be tuned by a simple method would be a breakthrough for fabricating ferroelectricity-based memories. Additionally, the system could be functionalized to exhibit magneto-electrical properties using external fields or strain that cause structural changes. For example, engineering heterostructures based on 2D oxides, hopefully, compatible with the highly advanced silicon technology \cite{2dmagnetoelecUlrich2022,ProgressGuan:2020bca}.

Within the family of two-dimensional materials, reported out-of-plane (OOP) 2D ferroelectrics are mostly based of two or more elements; however 2D elemental materials with OOP and in-plane ferroelectricity have also been reported, which suggests that crystal's symmetry as much as different electronegativities among others factors such as lattice stacking and ion-displacement are all but complementary tools to get sizeable polarization \cite{LeiLi2017,Yuan2019,GuanReview2020,Xiao2018,Yang2018}.

A bulk h-BN crystal has a layered structure, similar to graphite, and each BN monolayer displays a honeycomb lattice, with two atoms in the unit cell of boron (B) and nitrogen (N). Bulk h-BN is centrosymmetric; the stacking between layers is AA', which means one layer is rotated by 180$^o$. In that stacking, a B(N) atom sits atop the other layer's N(B) atom. We call this stacking antiparallel. In the parallel staking the top and bottom layers have the same orientation. In this configuration the AA stacking is energetically unfavorable but the AB or BA stacking has roughly the same formation energy as the AA'.
In the parallel stacking it is possible to find non-centrosymmetric stackings and obtain ferroelectric structures due to the polar nature of the monolayer. For instance, a bilayer AB is non-centrosymmetric and displays out-of-plane polarization. In AB stacking, one of the atoms lies on top of another atom in the second layer; the other atom lies in the center of other layer hexagon. The AB and BA stackings are related by mirror reflection with a net polarization pointing in opposite directions. Recently, several independent research groups found polarization in a h-BN bilayer up to room temperature \cite{Yasuda1458,Woods2021,Vizner2021}. 

Rotation between atomic monolayers has become an essential tool in the search for new properties; in the past has revealed unexpected surprises \cite{Morell2010,Cao2018a,Zheng2020}. When we rotate two layers for angles below one degree, a remarkable reconstruction occurs in the unit cell. Atomic and electronic reconstruction in 2D materials demonstrates the enormous influence that interlayer coupling has on the crystal structure and charge distribution, which directly affects the electrical polarization, as we will show below  \cite{twistronic2021}.

Indeed, in h-BN parallel stacking when one of the layers is rotated, regions with AB and BA stacking appear in the moir\'e cell. The size of those regions depends on the relative rotation angle and on the initial orientation of the bilayer structure \cite{Enaldiev2020,Vizner2021,delaBarrera2021}. 
Furthermore, in parallel stacking the relaxation process forces a reconstruction of the structures for very low angles, the unstable AA regions shrink, and the AB and BA regions increase their size. The structure is composed of alternating triangular regions with AB and BA stacking separated by incommensurate walls. The latter has been observed experimentally and obtained by molecular dynamics simulations \cite{Woods2021}.

We will show later that we can go from an AB to a BA stacking by moving a layer by a distance equal to the  B-N link. Moreover, electric polarization domains similar to those field-tunable ones seem e.g., in reference \cite{Vizner2021}, are predicted. Fortunately, in van der Waals-like materials, the super-lubricity between layers causes low energy barriers, around four meV for the h-BN bilayer, which allows an external electric field to switch from one stacking to the other and change the polarization direction. In fact, fields around 0.3 V/nm can be used to tune triangular-like polarization domains in rotated h-BN layers \cite{Vizner2021}.

This article studies a system composed of three h-BN layers in all possible independent configurations starting from a parallel stacking, as the antiparallel stacking configurations are centrosymmetric and lack OOP polarization \cite{Vizner2021}. We calculate the charge transfer between the layers, the electric polarization, and the energy barriers between the different configurations while migrating from one to another along the planar bonding direction. 
We propose that with three h-BN layers, one of them rotated, we can achieve different electric polarization checkerboards depending on the initial stacking and the rotated layer. The polarization values and the size of the regions depends on the initial stacking. 

The article is organized as follows: In section \ref{Metodos} we explain the methods used. All calculations of the unrotated trilayers: electronic structure, charge transfer and energy barriers were performed using Density Functional Theory (DFT) methods. The local charges were analyzed within a Bader scheme as well as relaxation of the rotated structures was performed with LAMMPS\cite{Plimpton1995}, and a tight-binding (TB) Hamiltonian was used to calculate the local charge and polarization in the rotated structures.
Section \ref{Bi_and_Trilaye} describes all the results for the AB bilayer and the different trilayer configurations.
Finally, in section \ref{Rotated} the impact that the rotation of one of the layers has on the polarization of the system is analyzed. 

\begin{figure*}[htb]
    \centering
    \includegraphics[width=0.95\linewidth]{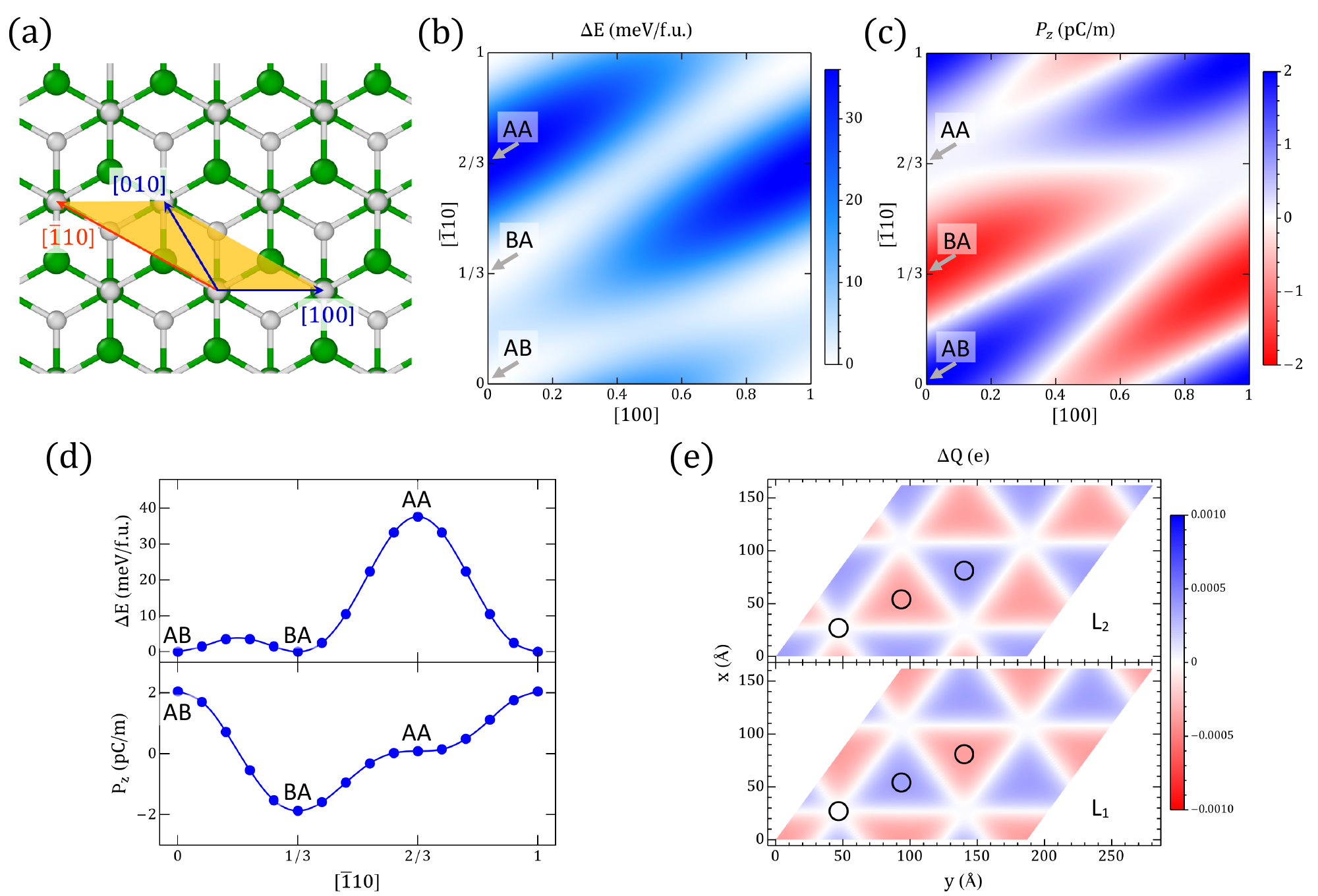}
    \caption{Top view of a bilayer AB unit cell in (a). The shaded parallelogram shows the area swept by the movement of the upper layer to calculate the energy barriers and polarization at each position. In (b) and (c) the energy barriers and out-of-plane polarization maps for different positions of the upper layer starting from an AB site, respectively. In each graphic, the horizontal axis correspond to the $[1 0 0]$ direction and the vertical axis to the $[\bar{1} 1 0]$. Density plots are obtained from interpolation over a $10 \times 15$ grid. The high symmetry points AA, AB and BA are marked. In (d) top and bottom panels the same as in (b),(c) but along the $[\bar{1} 1 0]$ direction. In (e) excess or defect of charge in a relaxed rotated bilayer structure. The black circles indicate regions where the polarization is calculated from a TB Hamiltonian.  }
    \label{fig:bilayer}
\end{figure*}

\section{Methods}
\label{Metodos}
Our calculations are performed by using DFT as implemented in the Vienna Ab-initio Simulations Package (VASP) \cite{Vasp1,Vasp2}. The generalized gradient approximation with the Perdew-Burke-Ernzerhof exchange-correlation functional \cite{PBE} and the projector augmented wave potentials \cite{Vasp2} are adopted. The kinetic energy cutoﬀ is set to be 520 eV, and a large vacuum space is set in the vertical direction so the nearest distance between two neighboring trilayer is greater than 15 \AA{}. For geometric optimization, the Brillouin zone is sampled with 12 × 12 × 1 and 24 × 24 × 1 k-points using the Monkhorst-pack scheme for relaxation and static calculations respectively, while the forces on all atoms are optimized to be less than 0.005 eV/\AA{} and the tolerance for the energy convergence is set to $10^{-5}$ eV. The PBE-D2 functional of Grimme \cite{Grimme2006} is used to take into account dispersive forces, and the Berry-phase method \cite{Vanderbilt1993} is employed to evaluate crystalline polarization. Electric polarization switching paths connecting different trilayer's stackings are calculated using the nudged elastic band (NEB) method\cite{Henkelman2000} with forces converged below 0.01 eV/\AA{}. We also used a classical atomistic model implemented in the LAMMPS code \cite{Plimpton1995} to relax the rotated structures, which are usually too large to be handle by DFT relaxations routines in reasonable times. The reconstruction of the cell impacts the crystal structure and consequently the electronic properties of the system. The relaxed structure is used to calculate the charge from a TB-type Hamiltonian. The electric polarization is also calculated using a Bader approximation \cite{Bader1985} and the dipolar correction \cite{dipolecorrecPRB.51.4014} for the charge distribution, while the differential charge distribution and transfer was performed with Vesta \cite{Momma:db5098} and Mathematica \cite{Mathematica}.\\

We use a TB model to analyze in the different structures how the charge is distributed between and within the layers. Although DFT methods can also do this, the study of twisted structures would require extremely large supercells which are still beyond typical computational capabilities. However, TB modeling is able to capture a majority the electronic effects in these systems to a cheaper computational cost while providing us with an insight into the settled but meaningful trends that arise from the comparison between minimalist and ab-initio calculations. 
  
The TB model includes a first-neighbor hopping within the layer and a distance-dependent hopping between the layers. We employed the parameters suggested in Ref. \cite{Guinea2021}.   
We set the onsite energy of Boron and Nitrogen sites $E_{B,N} = \pm 4$ eV, the intralayer nearest neighbor hopping $t = 2.33$ eV, the hopping between layers depends on the atoms, in the direct position $t_{BB}$ = 0.7 eV, $t_{NN}$ = 0.15 eV, and $t_{NB} = 0.3$ eV, the distance depending hopping follows a standard exponential decay $t_{XY}(r)=t_{XY} \exp[-\alpha(r-d)]$ \cite{Morell2011,Guinea2021} where $\alpha=4.4$ and $d=3.35$ \AA{}.

In Appendix \ref{Bands_app} we show the calculated bands of two h-BN trilayers projected on B and N orbitals. The highest valence and lowest conduction bands are  mostly composed of electrons in $p_z$ orbitals of the two atoms, validating the use of a TB model with only one orbital.

From the TB model described above we calculate the charge transfer between layers.  In two-dimensional systems with vertical polarization we do not have the ambiguity in the definition of the unit cell for the polarization calculation.  As we will show below, the polarization calculated from the charges resembles closely the main properties obtained by the Berry's phase calculations. Moreover, such charges analysis allows us to discriminate between possible local stackings in rotated trilayers where it is difficult to distinguish between them with the naked eye.

\section{Bilayer and Trilayer}
\label{Bi_and_Trilaye}
\subsection{Bilayer h-BN}
\label{bilayer_hBN}
\subsubsection{Polarization and energy barriers}
Before looking at the case of trilayer h-BN, we calculate the bilayer structure to setting up our calculations and compare the polarization results with the three-layer system. 

In Figure \ref{fig:bilayer}(b) and (c), we show the maps of the energy barrier  and the OOP polarization for different positions of the upper layer, keeping the lower layer fixed. The starting position is an AB stack. The horizontal axis corresponds to translations along with the B-B bond and the vertical axis along the B-N bond. The map interpolates $10 \times 15$ points; the system was not structurally relaxed in each position except the AB. However, the results agree with previous calculations\cite{Lebedev2016}. The lowest energy positions correspond to stacks AB and BA, while AA has a 40 meV barrier. 

The polarization map shows vast zones with non-zero polarization and some fringes where symmetry prohibits net Polarization. The maximum value of polarization corresponds to sites AB(BA) and it is $\pm 2$ pC/m similar to a previous result \cite{LeiLi2017}.

In Figure \ref{fig:bilayer}(d) we show the energy barriers and the polarization path along the $[\bar{1} 1 0]$ direction. This direction is peculiar because we can move from a stack AB to a BA, which have polarization pointing in opposite directions, and later to AA; each stack is separated from the previous one by a distance equal to the B-N bond. The direction joining the AB and BA sites faces an energy barrier of only 4 meV per formula unit. In those two sites the absolute value of polarization is maximum and point in opposite directions. The AA site has the highest energy barrier and the polarization is zero as expected from its P\={6}m2 symmetry.
\\

\subsubsection{Bader charge analysis}
We proceed now to estimate the charge transfer between the layers in the polarized AB phase,  to do that we performed a Bader charge analysis as implemented in the Bader charge code\cite{BaderSite}. This method encloses the atoms in surfaces where the charge density is minimized, and later the charge inside those surfaces is obtained.
The mean Bader charges of B and N of each layer, as well as their respective effective charges (calculated as the difference between the number of valence electrons $Z_{val}$ from the pseudo-potentials and the Bader charges \cite{costa-amaral_role_2021}) are presented in Table \ref{tab:hbn_bi_bader}. 

We carry out a classical calculation of the polarization of the system considering the effective charges as points. Using a distance between the layers $d=3.1$ \AA{} we obtain a value  $P_z=2.74$ pC/m, fairly consistent with the value obtained by ab initio calculations that use modern polarization theory. On the other hand,  Bader charge differences $\Delta Q_B$ with respect to the monolayer  (see Table \ref{tab:hbn_bi_bader}) reveal an unequal distribution of charges between the layers. This suggests that the main contribution to the OOP polarization comes from a small charge transfer from one layer to another.


\begin{table}
\centering
\caption{Mean Bader ($Q_{Bader}$) and effective charges ($Q_{eff}$) of B and N in the monolayer and in the lower (${L_1}$) and upper (${L_2}$) layers in AB-stacked bilayer h-BN. $\Delta Q_{B}$ is the difference in $Q_{Bader}$ between bilayer and monolayer h-BN.}
\label{tab:hbn_bi_bader}
\begin{tabular*}{\tblwidth}{lcccc} 
\toprule
Monolayer              & \multicolumn{2}{c}{B}       & \multicolumn{2}{c}{N}         \\ 
$Z_{val}$                & \multicolumn{2}{c}{3}       & \multicolumn{2}{c}{5}         \\
$Q_B$                  & \multicolumn{2}{c}{0.78812} & \multicolumn{2}{c}{7.21188}   \\
$Q_{eff}$                & \multicolumn{2}{c}{2.21188} & \multicolumn{2}{c}{-2.21188}  \\ 
\midrule
Bilayer AB             & B ($L_1$) & N ($L_1$)        & B ($L_2$) & N ($L_2$)          \\ 
$Q_B$                  & 0.78673   & 7.21629          & 0.78743   & 7.20955            \\
$Q_{eff}$                & 2.21327   & -2.21629         & 2.21257   & -2.20955           \\
$\Delta Q_B$           & -0.0014   & 0.00441          & -0.00069  & -0.00233           \\ 
$\Delta Q_B$ per layer & \multicolumn{2}{c}{0.00301} & \multicolumn{2}{c}{-0.00302}  \\
\bottomrule
\end{tabular*}
\end{table}

\subsection{Trilayer h-BN}

\subsubsection{Structures, Polarization and energy barriers}
With three h-BN layers there are at least nine high symmetry configurations to consider. 
In table \ref{tab:path} we show how we can go from ABA stacking to any other configuration by shifting one of the layers. The value $m$ is associated with the middle layer and $u$ with the upper layer, they take  fractional values and reflects a shift of each layer along the diagonal $[\bar{1} 1 0]$. Among those configurations, there are three stackings that are related to an equivalent configuration by an OOP mirror reflection.

\begin{table}
\centering
\caption{h-BN trilayer stacking configurations.}
\label{tab:path}
\begin{tabular*}{\tblwidth}{@{}CCCC@{}} 
\toprule
 &  \multicolumn{3}{@{}C@{}}{$u$}           \\ 
\cline{2-4}
$m$    & 0   & 1/3                & 2/3                 \\ 
\midrule
0   & ABA & ABB                & ABC                 \\ 
\midrule
\multirow{2}{*}{1/3} & \multirow{2}{*}{ACA} & ACB & \multirow{2}{*}{ACC}                 \\ 
 &  & (inverted ABC) &                  \\ 
\midrule
\multirow{2}{*}{2/3} & \multirow{2}{*}{AAA} & AAB & AAC \\
  &  & (inverted ACC) & (inverted ABB)  \\
\bottomrule
\end{tabular*}
\end{table}

Table \ref{tab:hbn_tri_pz} contains, for each structure, the exact values of the polarization with and without including the dipole correction, the energy of the system with respect to the most stable configuration and the distance between the layers, which varies depending on the stacking; it is minimal in AB stacking but reaches its highest value in AA as expected. 

The four lowest energy configurations have nearly the same energy, with either zero or maximal OOP polarization. There are four other stackings 16 meV/f.u. higher in energy and the unstable AAA stacking with a 32 meV/f.u. barrier.

The absolute value of polarization is maximum at ABC and ACB configurations with values of $\pm 5$ pC/m, pointing in opposite directions. This value is $\sim$ 2.5 times larger than bilayer. 
The net polarization is zero in the ABA, ACA and AAA stackings in accordance with their P\={6}m2 OOP mirror symmetry. The six stackings with non-zero polarization belong to P3m1, where such symmetry is broken.

\begin{table}
\caption{Out-of-plane electric polarization ($P_z$), in parentheses including the dipole corrected value, configuration energy relative to the lowest energy configuration (ACA) ($\Delta E$), interlayer distances between the lowest and middle layers ($d_{L_1, L_2}$) and between middle and upper layers ($d_{L_2, L_3}$), for the nine h-BN trilayer configurations. \label{tab:hbn_tri_pz}}
\begin{tabular*}{\tblwidth}{@{}CCCCC@{}} 
\toprule
    & $P_z$         & $\Delta E$ & $d_{L_1, L_2}$ & $d_{L_2, L_3}$  \\
    & (pC/m)        & (meV/f.u.) & (\AA{})        & (\AA{})         \\ 
\midrule
ABA & 0.0 (0.0)     & 0.23       & 3.07341    & 3.07375     \\ 
ABB & 2.48 (1.63)   & 15.96      & 3.07783    & 3.39648     \\ 
ABC & 4.97 (3.28)   & 0.24       & 3.07452    & 3.07713     \\ 
ACA & 0.0 (0.0)     & 0.0        & 3.07425    & 3.07386     \\ 
ACB & -5.01 (-3.29) & 0.25       & 3.07566    & 3.07241     \\ 
ACC & -2.48 (-1.62) & 15.96      & 3.07443    & 3.41738     \\ 
AAA & 0.0 (0.0)     & 31.77      & 3.41350    & 3.41354     \\ 
AAB & 2.45 (1.62)   & 15.95      & 3.41142    & 3.07919     \\ 
AAC & -2.40 (-1.62) & 15.97      & 3.41527    & 3.08198     \\
\bottomrule
\end{tabular*}
\end{table}

\begin{figure}[ht]
    \centering
    \includegraphics[width=0.95\linewidth]{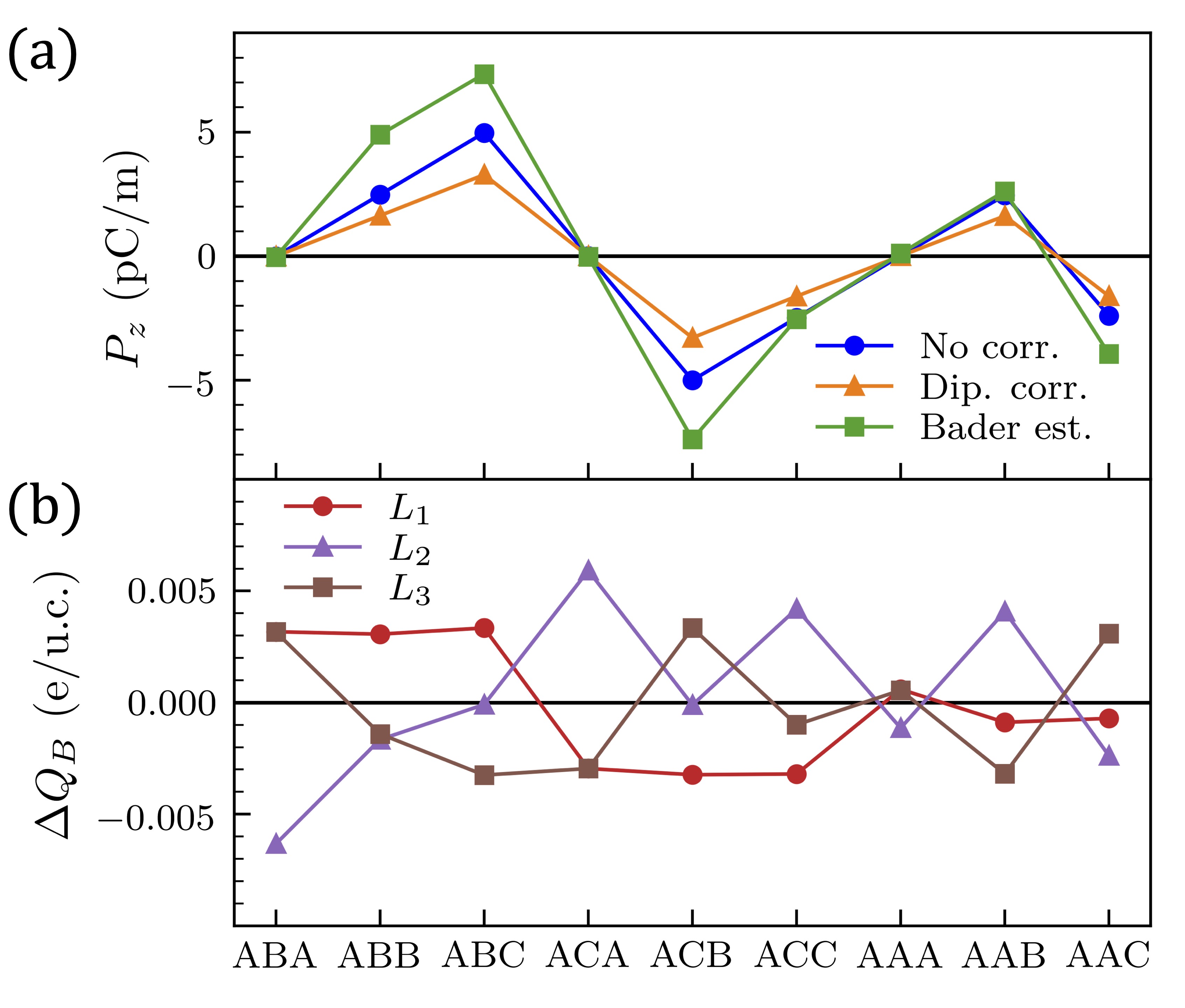}
    \caption{(a) Out-of-plane electric polarization $P_z$ for each of the h-BN trilayer stacking configurations and (b) Bader charge excess/deficiency of each layer with respect to the monolayer h-BN ($\Delta Q_B$). The electric polarization is obtained from Berry's phase calculations with/without dipole corrections, as well as estimated from a point-charge model considering the Bader charges of the ions in their respective stacking configuration. }
    \label{fig:pz_bader_trilayer}
\end{figure}

In Fig. \ref{fig:pz_bader_trilayer} (a) we plot the polarization values from Table \ref{tab:hbn_tri_pz} plus the polarization obtained from the Bader charges analysis. The latter overestimate the DFT calculation as it does not consider the spatial distribution of the charge density, but correctly predicts the sign.  
Dipolar corrections not only test the default charging aspects of our slabs but also allow us to value the differences between polarization calculated from Bader charge and Berry's phase theory. 
Although a simulation of the experimental realization is not our primary target, but rather to get a further insight into the polarization trends and possibilities, it is worth to note that the values given by the dipolar correction puts our predictions within a range of plausible experimental values. DFT calculations in 2D systems require a vacuum in z to avoid the interaction of the layers due 
to the periodicity requirement. To comply with the periodicity of the potential, if no dipole correction is applied, a finite displacement field is created in the vacuum region, which can lead to extra charge transfer between the layers\cite{Ferreira2021,dipole_HE}. Dipole corrections remediates this and in our particular case are also showing that in the key configurations displaying large
electric polarizations all the models presented here predict fairly the same trends, given that Bader's overestimation would be corrected similarly to lower values.

Fig. \ref{fig:pz_bader_trilayer}(b) shows the excess or deficiency of charge in each layer for each of the configurations, with respect to isolated monolayers. The values are a reflection of the symmetry of the system and can be used to very quickly obtain an estimate of the polarization in each of the nine configurations. In the ABA, ACA and AAA stacks the charge transfer is symmetrical with respect to the middle layer. The latter receives/gives the charge given/received by the outer layers.

In the ABC and ACB stacks, where the polarization is maximum, the excess/defect charge in the middle layer is almost zero, a transfer occurs between the outer layers mediated by the middle layer. This causes the polarization value to be maximized. Additionally, these values can be used to predict the stacking between layers in different regions of rotated layers, as we will see below.

\begin{figure*}[htbp]
    \centering
    \includegraphics[width=0.95\linewidth]{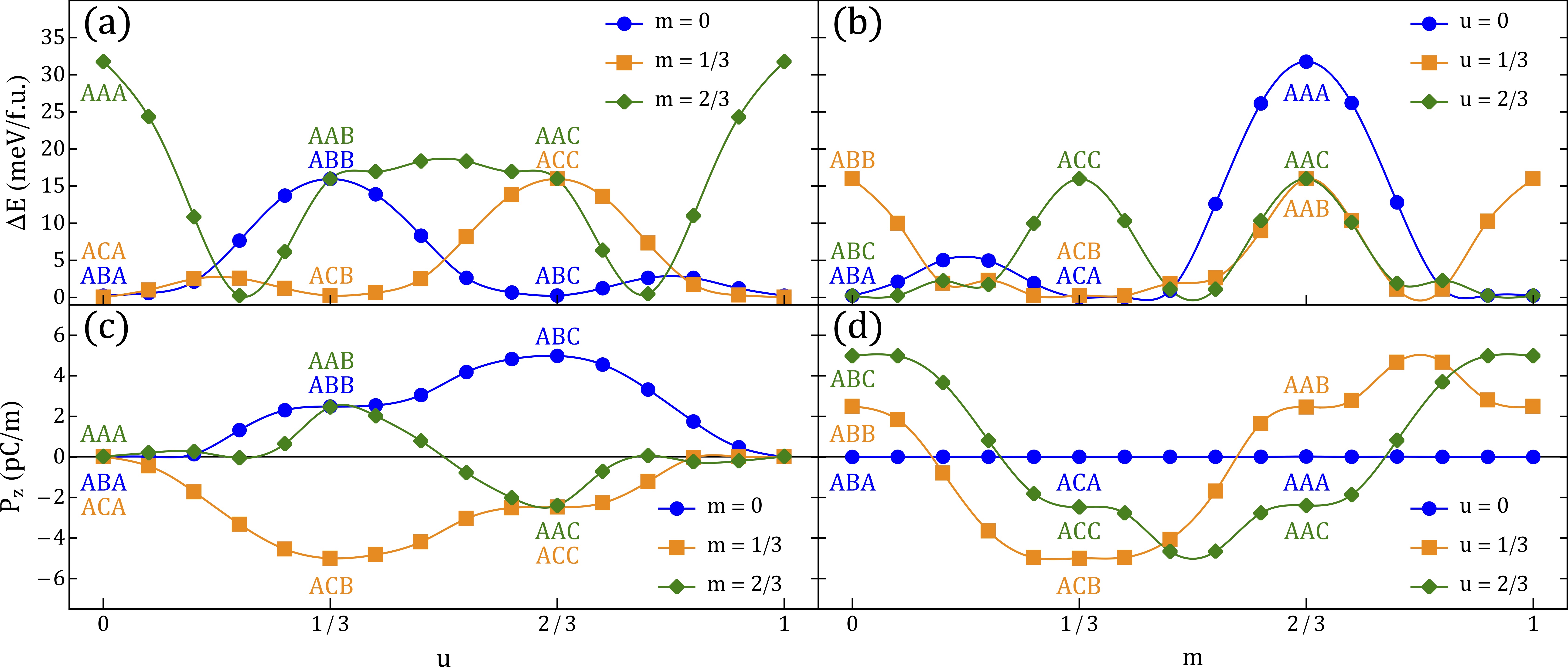}
    \caption{Energy barriers (a,b) and out-of-plane electric polarization (c,d) obtained from NEB relaxation. In the left(right) panels we keep the middle(upper) layer fixed in three different positions and shift the upper(middle) layer. The stackings for different values of m and u are described in Table \ref{tab:path}. }
    \label{fig:EnergyBarriersTrilayer}
\end{figure*}
\subsubsection{Energy barriers and pathways}
Fig. \ref{fig:EnergyBarriersTrilayer} shows the energy barriers (top panels) and polarization (bottom panels) along different paths. The left (right) panels show for three fixed positions of the middle (top) layer the change of both quantities as the top (middle) layer is shifted. 
Different paths can be observed, some of them are energetically unfavorable due to their large barriers or stacking instabilities, but we will concentrate on those that we consider can be more feasible to obtain experimentally, for example a shift of one layer under the application of an external electric field.

The barrier between the ABA and ABC stacks is only 4 meV (blue line, panel (a)) and in the first case the vertical polarization is zero and is maximum in ABC as can be seen in (c). 
The ACA-ACB path is the complementary case to the previous one (orange lines) it is also interesting because the barrier is also small and in this case the polarization goes from zero to the smallest possible negative value.
In both paths the 4 meV barriers resemble the AB-BA switching in h-BN bilayer, with the two upper layers acting as an effective AB-BA switching while the bottom layer offsets the polarization.

The landscape in Fig. \ref{fig:EnergyBarriersTrilayer} suggests the existence of several paths for which the polarization change ranges within $[0,5]$ pC/m for one-layer solo movement. The low-energy barriers below 5 meV, i.e., from ACA to ACB and from ABC to ABA, stand as promising paths as the polarization change is due to a paraelectric-like to ferroelectric state transition. The latter means that using an electric field to assist the layer's movement would not waste too much energy along the whole path in terms of polarization switching before complete polarization. On the other hand, we attain the largest polarization change if we could move two complementary layers step-wise, from ABC to ACB; the polarization changes would be the largest, and the barriers are below 5 meV at each step. 

Moreover, the three layers system offers insights into how meta-stable states could also be used for electrical switching, as displayed by the AAB to AAC path. The energy cost for the system to travel this path is slight as we have a plateau-like transition; however, a fairly large polarization reversion is reached, also having the possibility of, at the endpoints, either a fast transition to the energy minimum, with almost null polarization or continue moving just one layer while maintaining the polarization.
These characteristics are unique to the three-layer systems and are an addition to the already advantage of having 20-30\% larger polarization as compared to the bilayer.

\section{Rotated layers}
\label{Rotated}
Let us now analyze a system in which, in a reasonably controlled way, regions with different polarization can be obtained in the same crystal. Several experimental works have shown that when two layers of a polar structure are rotated, regions with opposite polarization are formed.  This is due to the fact that three very distinctive regions appear in the structure, a zone with AA stacking, another zone with AB and a third with BA. As we analyzed above, in Section \ref{bilayer_hBN}, the polarization is opposite in the AB and BA stacks and zero in the AA. Additionally, due to the nature of the atoms composing the monolayer and the energy of the different possible stackings, the crystal undergoes a reconstruction minimizing the unstable AA zones and increasing the size of the AB(BA) zones.

Indeed, we performed a relaxation process to a rotated h-BN bilayer structure at an angle close to 1\textdegree{} using the previously described LAMMPS code, and after the relaxation the structure undergoes a reconstruction and the size of AB/BA zones are increased. To corroborate this we perform a calculation of the charge transfer between the layers using a TB Hamiltonian and in Fig. \ref{fig:bilayer}(d) we plot a map of the excess/defect charge in each layer. Alternating triangular regions with positive/negative charge, i.e. regions with opposite polarization, can be observed, our results are in perfect agreement with what was obtained previously \cite{Guinea2021, Woods2021}. 

\subsection{Twisted trilayer}

When we have three layers and one of them is rotated, it is difficult to visually discern the stacking of the regions. To solve this problem, we decided to calculate the charge and polarization locally in different regions. The stacking in different regions of the moiré unit cell must be a combination of at least three of the nine stackings we studied earlier. Which ones we obtain depends mainly on the rotated layer and the initial stacking of the system. 

We describe now the procedure to overcome that problem: First, the rotated structure is submitted to a relaxation process as explained above. Later, using the relaxed cell we proceed to calculate the charge transfer between the layers and analyse locally the behaviour of the charge. For this we concentrate on three zones along the main diagonal of the moiré cell, that due to the symmetry of the system have different behaviour. The three points are separated by one third of the main diagonal size. We calculate then at an adjustable radius around these three points the charge in each layer and the resulting polarization.

\begin{figure*}[ht]
    \centering
    \includegraphics[width=0.95\linewidth]{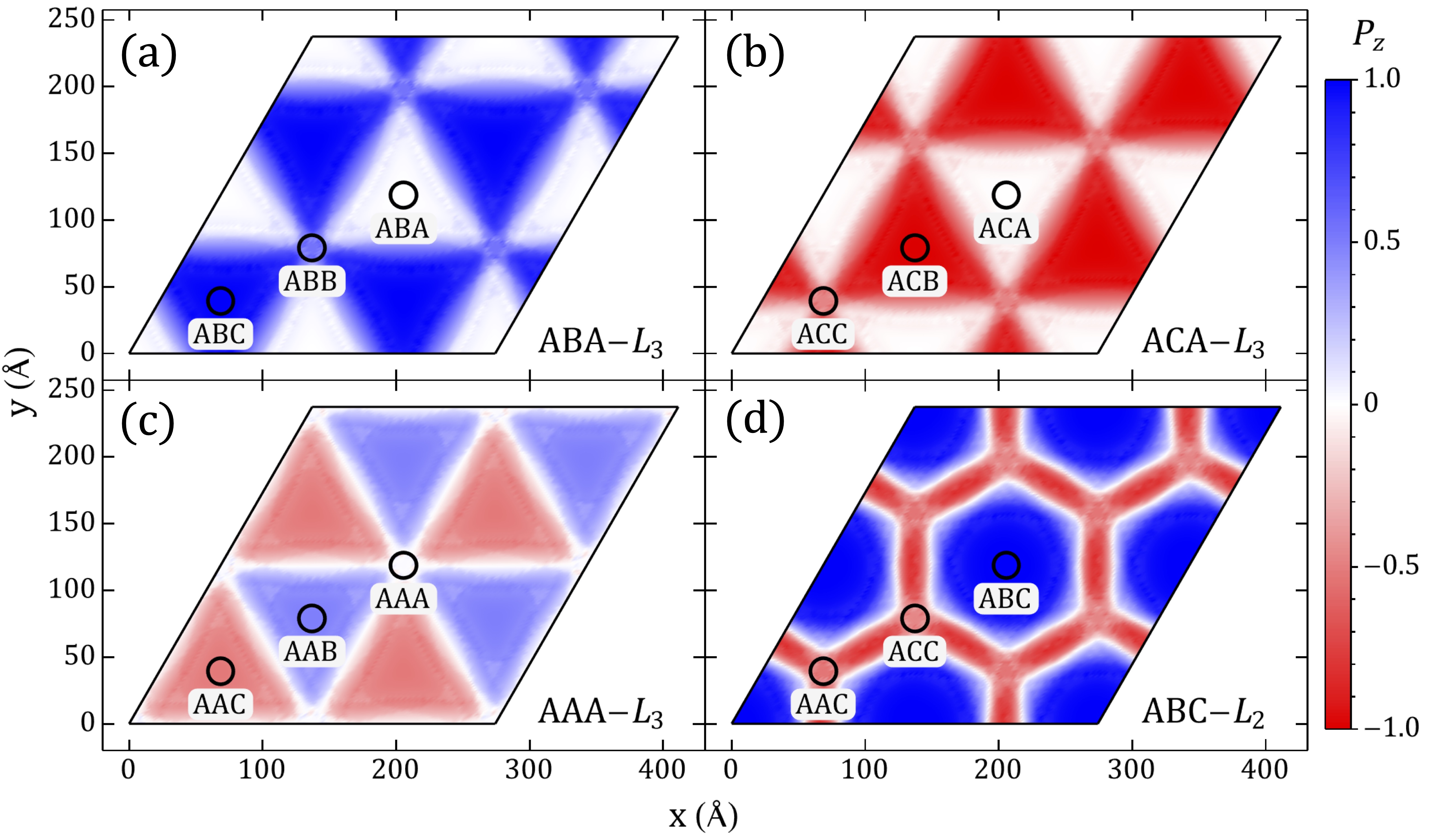}
    \caption{\label{fig:trilayer-twisted-pz} Twisted h-BN trilayer out-of-plane polarization for ABA, ACA and AAA with top-layer rotation (labeled ABA-$L_3$, ACA-$L_3$, AAA-$L_3$ respectively), and ABC with middle-layer rotation (labeled ABC-$L_2$). The circles show the regions chosen to calculate the charge in each layer and the polarization shown in Table \ref{tab:hbn-twisted-cargas-stack}. Polarization values are normalized to the value at the ABC region.}
    \label{fig:Twisted}
\end{figure*}

Let us now consider four possible scenarios that produce regions with different polarization. In the first three from an ABA, ACA, or AAA starting configuration, we rotate the third layer, and in the fourth from an ABC we rotate the middle layer. Fig. \ref{fig:trilayer-twisted-pz} shows the polarization map for these four configurations, normalized to the maximum absolute values in the ABC and ACB stackings. In Appendix \ref{Charge_app} we analyze the charge redistribution between the layers for all these four configurations.

In Fig. \ref{fig:trilayer-twisted-pz}(a), starting from an ABA, we obtain triangular regions with "positive" polarization, with  ABC stacking, alternating with triangular regions with zero polarization, ABA stacking. The size of the ABB zone is considerably smaller than the two triangular regions.
The result is the same if the starting stacking is ABC and the third layer is rotated. In this case only the location of the positive and zero polarised zones changes. Now, in Fig. \ref{fig:trilayer-twisted-pz}(b) we start from an ACA stacking and obtain a similar pattern but with opposite polarization due to an ACB stacking in the red triangular areas.

The configuration shown in Fig. \ref{fig:trilayer-twisted-pz}(c) might be challenging to obtain and we show it for completeness. Because the first two layers would be AA-stacked, which, as we have shown, is energetically unstable, we had to force the first two layers not to relax to obtain this structure. This configuration is analogous to the previously shown h-BN bilayer with alternating positive and negative polarised zones.

Finally, the pattern in the fourth configuration shown in Fig. \ref{fig:trilayer-twisted-pz}(d) is completely different. Starting from an ABC stacking we rotate the middle layer and obtain a large hexagonal region with local ABC stacking surrounded by walls with opposite polarization of alternating AAC and ACC alignments at the vertices.

In Fig. \ref{fig:twisted_charges_perlayer}(d) from \ref{Charge_app}  we show the map of the charge in each of the three layers. There are areas of excess and deficiency of charge in layers 1 and 3, but the regions are not symmetrical. The middle layer instead shows large areas with little change in the charge and with maximum polarization. The other two ACC and AAC zones in turn are smaller in size due to the direct coupling between two layers, akin to the shrinking of the unstable AA regions in the bilayer case.

Interestingly, using the $L_2$ layer allows us to induce polarization domain walls, which enclose polarized hexagonal dots whose size can be tuned, and also the  
width of such walls.
This might be promising because the switching of the polarization in a specific region would most likely mean a domino effect in the checkboard, as seen in previous work \cite{Zhang2022}. 
The reversal of the polarisation walls is different from the center points in terms of energy, as the walls are continuously connected, which promotes isolation of the points that could be tuned through angle and initial stacking.

In Table \ref{tab:hbn-twisted-cargas-stack} we show for two twisted configurations the excess or defect of charge in each layer in terms of the electron charge and the resulting polarization obtained from a classical calculation, for every of the three regions marked with circles in Fig. \ref{fig:trilayer-twisted-pz}. The polarization values are normalized to the ABC maximum value. For the configuration shown in Fig. \ref{fig:trilayer-twisted-pz}(d) we can see that the polarization at the vertices is almost one tenth smaller in magnitude than in the ABC region.

It is also possible to produce configurations without polarized regions. For example if we take an ABA stacking and rotate the middle layer the polarization is zero in the three regions mentioned. This creates ABA, ACA or AAA local stackings which by symmetry have no polarization. 

We point out that although these results were obtained with a TB Hamiltonian, in all regions the layer charges are consistent with the analysis performed by first-principles methods and with the values obtained from the Bader charge method. Compare for example the results for the local ABC stacking of Table \ref{tab:hbn-twisted-cargas-stack} with Fig. \ref{fig:pz_bader_trilayer}(b).

\begin{table}
\centering
\caption{Excess/Defect of charges in each layer $C_{L_{i}}$ per \AA{}$^{2}$, out-of-plane polarization normalized to ABC value and local stacking configuration at the indicated sites for ABA with top-layer rotation and ABC with middle-layer rotation.}
\label{tab:hbn-twisted-cargas-stack}
\begin{tabular*}{\tblwidth}{@{}CCCCCCC@{}} 
\toprule
\multicolumn{1}{c}{}      & Site & $C_{L_1}$ & $C_{L_2}$ & $C_{L_3}$ & $P_z$ & Stacking  \\ 
\midrule
\multirow{3}{*}{ABA-$L_3$} & 0    & -0.21   & 0.41    & -0.20   & 0     & ABA       \\ 
                          & 1/3  & -0.21   & 0.016   & 0.20    & 1     & ABC       \\ 
                          & 2/3  & -0.21   & 0.20    & -0.002  & 0.47  & ABB       \\ 
\midrule
\multirow{3}{*}{ABC-$L_2$} & 0    & -0.13   & 0.0001  & 0.13    & 1     & ABC       \\ 
                          & 1/3  & -0.001  & 0.12    & -0.11   & -0.12 & AAC       \\ 
                          & 2/3  & 0.10    & -0.11   & 0.01    & -0.12 & ACC       \\
\bottomrule
\end{tabular*}
\end{table}

These simulations were performed for an angle of 1.05 degrees, the behavior is similar for smaller angles although the area of the main zones increases and the AAA zones become smaller.

In graphene trilayers it has been possible to change the stacking from ABA to ABC and back using a double gate \cite{Li2020}.
It is not unreasonable to think of a three-layer h-BN system, one of them rotated, with ABA/ABC zones and with external control of stacking and thus polarization. Furthermore, an external electric field could modify the stacking of the system \cite{Weston2022}.

From an experimental point of view, the use of trilayers is also appropriate, as any functionalization of a multilayer would most likely require one of the layers to be constrained by the substrate, leaving us with one or two layers we could control to generate a specific polarization pattern. Figures \ref{fig:EnergyBarriersTrilayer} and \ref{fig:Twisted} illustrate the various ways in which different polarization patterns can be achieved.

\section{Conclusions}

In summary, we have studied three h-BN layers and assessed their use as ferroelectrics. We first studied for all possible stackings, the polarization of each one and the energy barriers between each crystalline phase using DFT calculations and then we proposed the rotation of one of the layers as a mechanism to obtain from an initial ABA/ABC configuration alternating triangular zones with and without polarization. The polarization could be reversed with external mechanisms such as a gate or an electric field in compliance with the basic principle of information storage. Our results can be extended to other polar multilayers such as transition metal dichalcogenides.

\section*{Acknowledgements}
ECE, ESM and JMF acknowledge financial support from FONDECYT Regular 1221301 (Chile) and USM-DGIIE Proyecto Investigaci\'on PI-LIR-2021-100

\section*{Appendices}
\appendix
\setcounter{section}{0}

\section{Charge density redistribution}
\label{Charge_app}

\begin{figure*}[htbp]
    \centering
    \includegraphics[width=0.95\linewidth]{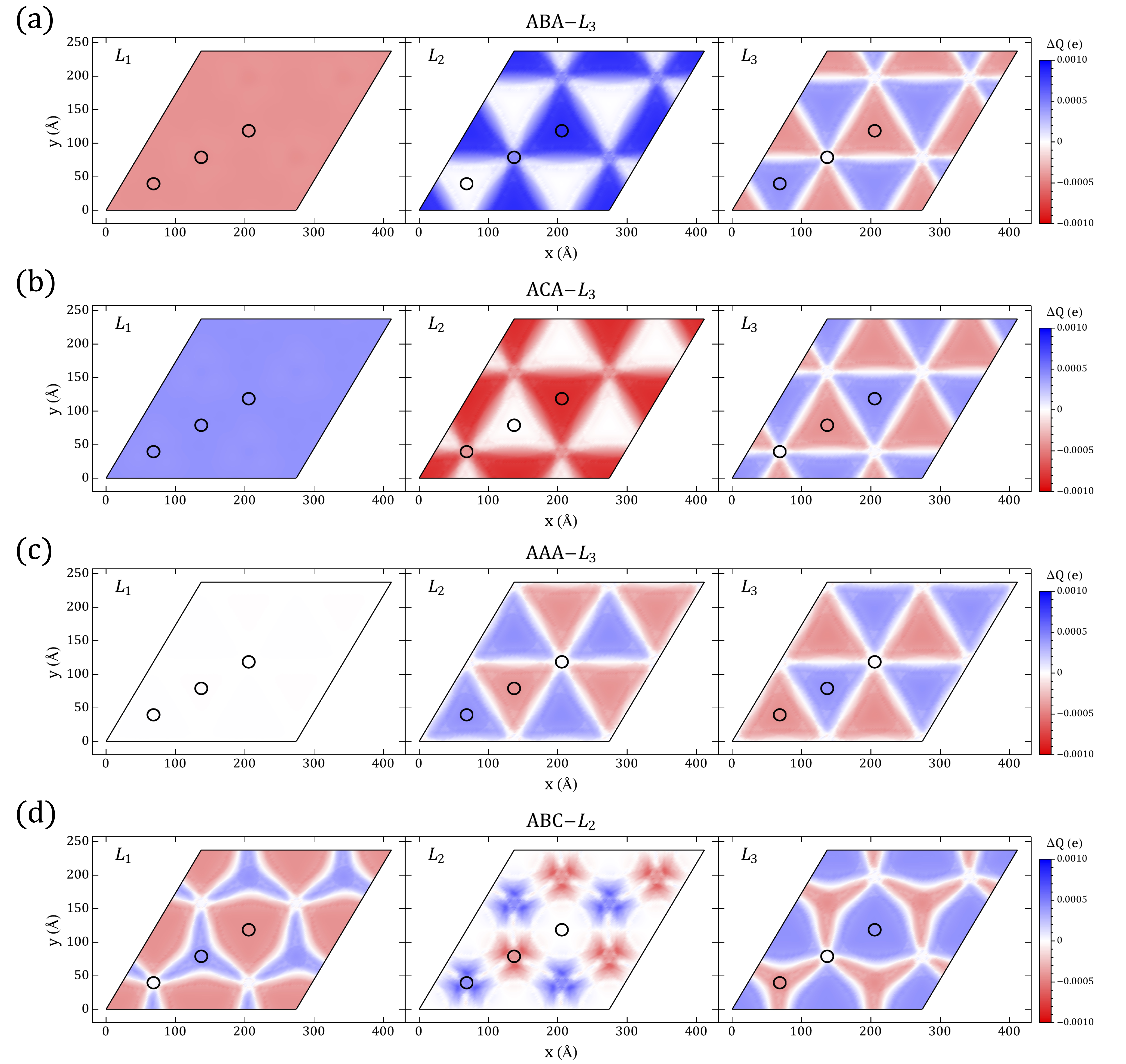}
    \caption{Twisted h-BN trilayer charge density per layer for (a) ABA, (b) ACA and (c) AAA with top-layer rotations and (d) ABC with middle-layer rotation, corresponding to the twisted configurations of Fig. \ref{fig:trilayer-twisted-pz}  of the main text. }
    \label{fig:twisted_charges_perlayer}
\end{figure*}

We have calculated using a TB model the charge redistribution in each rotated structure. In Fig. \ref{fig:twisted_charges_perlayer}, we show the excess/defect of charge per layer for the four configurations mentioned in the main text. In (a) we start from an ABA structure with a top layer rotated by an angle of 1.05 degrees. The red colour indicates loss of charge and the blue colour excess or gain. 

It can be seen that the lower layer loses charge while layers 2 and 3 have a behavior similar to that of the rotated bilayer, i.e. they have areas where charge is gained or lost with a triangular pattern. There are small differences in this case, the middle layer has areas where charge is gained and others where there is practically no change, the white triangles. The upper layer does have areas of excess and deficit of charge. 

The analysis of the three zones mentioned indicates that in (0;0) the polarization is zero so the stacking would be ABA, in the site (1/3;1/3) the polarization is maximum consistent with an ABC stacking and in the third site, which is smaller in diameter, have an ABB stacking, the polarization is non-zero. The values of polarization in each region for configurations (a) and (d) are listed in Table \ref{tab:hbn-twisted-cargas-stack} Note that due to plotting several unit cell the (0,0) points are the middle circles in all figures.

Notice that in \ref{fig:twisted_charges_perlayer} (a), (b) and (c) the stacking between layers one and two is AB, AC and AA respectively, there is no rotation between the layers, so the pattern is the same as in the primitive cell.  In the first case layer one loses charge, in the second case it gains charge and in the last case due to the AA stacking there is no charge transfer between the layers due to symmetry.

\section{Band structure of h-BN trilayers}
\label{Bands_app}

\begin{figure*}[htbp]
    \centering
    \includegraphics[width=0.9\linewidth]{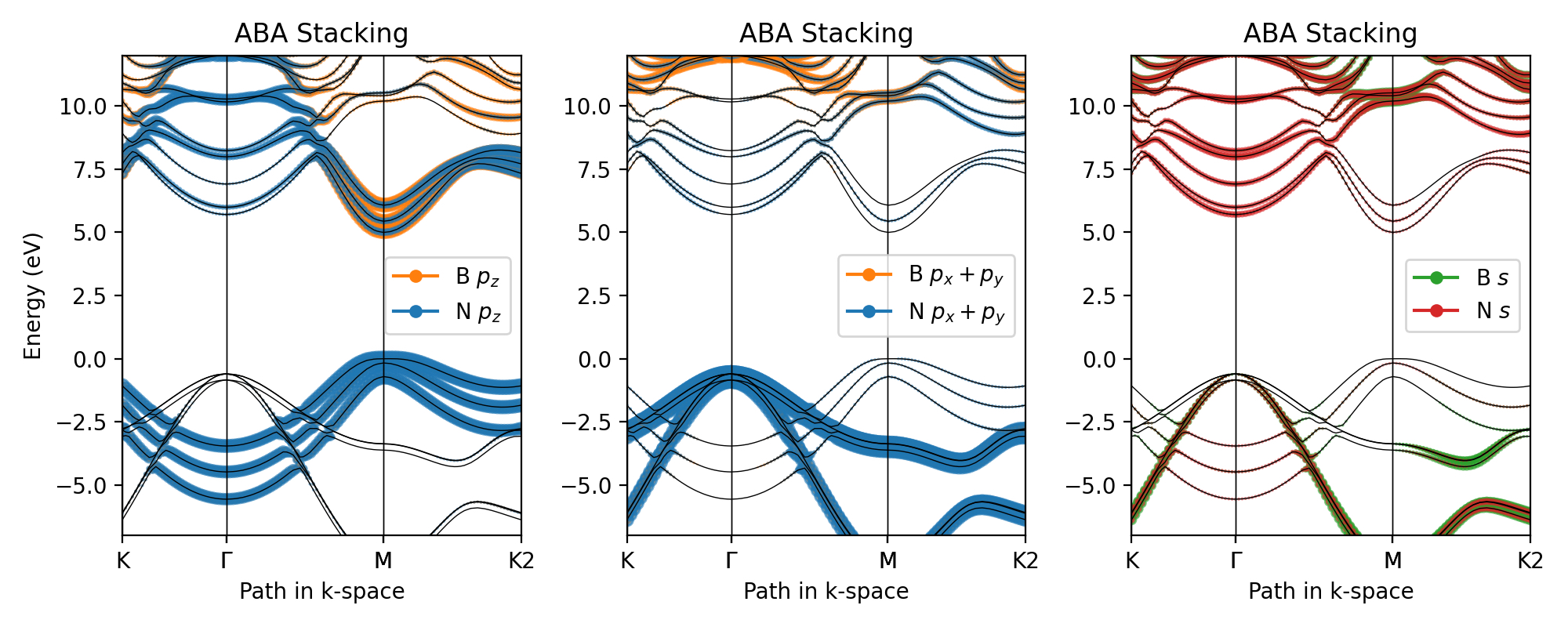}
    \includegraphics[width=0.9\linewidth]{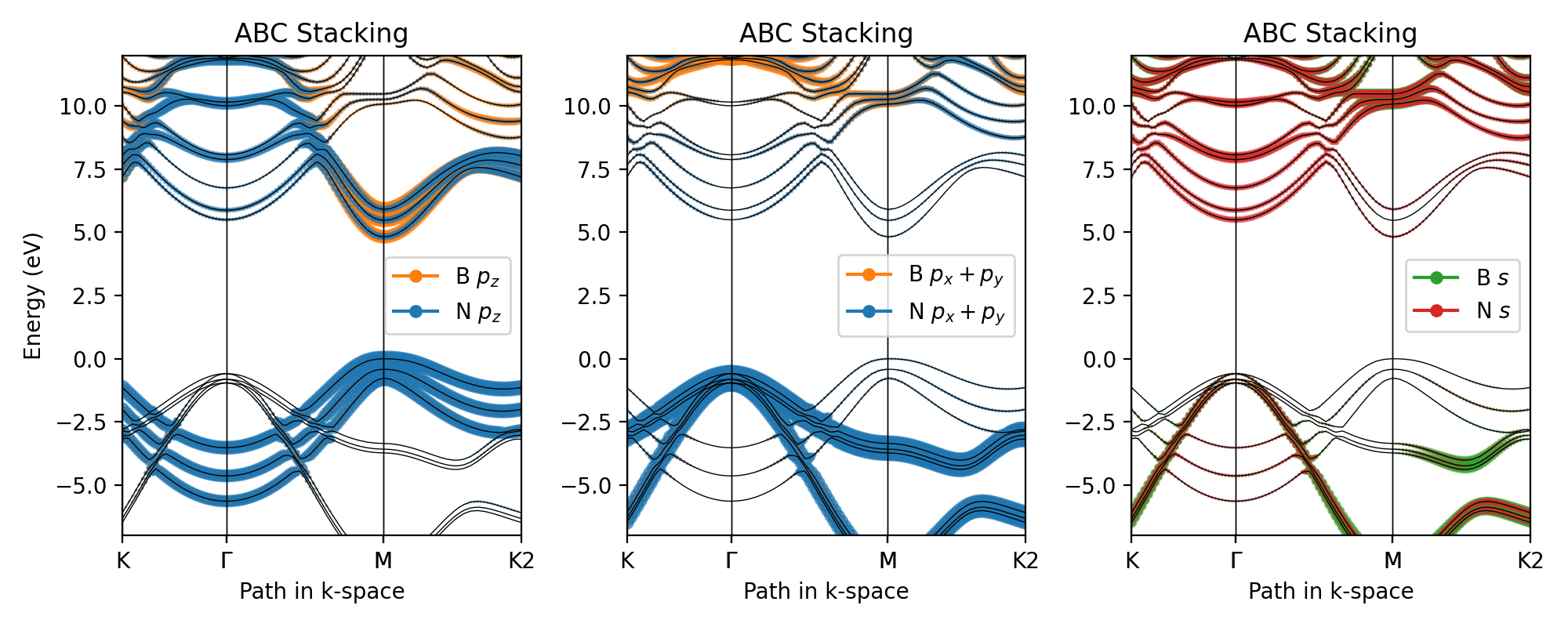}
    \caption{Orbital-projected band structure of ABA (top) and ABC (bottom) trilayer h-BN. Each panel shows the projected weight of the bands on the orbitals listed in the inset frame.}
    \label{fig:bands_orbital}
\end{figure*}

In Figure \ref{fig:bands_orbital} we show the projected bands of ABA and ABC trilayer h-BN on orbitals $p_z$, $p_x+p_z$ and $s$ of each atom. The calculations were done with VASP.  \\
The figures reveal that the gap is located at the M point and the highest valence and lowest conduction bands are better characterized by B and N $p_z$ orbitals. The weight of the orbitals $p_x+p_y$ is more significant around the $\Gamma$ point, which is lower by more than 250 meV from the Fermi level. The latter justifies that the use of only $p_z$ orbitals in the TB model gives a good agreement with DFT calculations on the charge redistribution. The $s$ orbitals of each atoms have a minor contribution on the states near the Fermi level.

\bibliographystyle{model1-num-names}

\end{document}